\documentclass[twocolumn]{aastex631}
\usepackage{mathrsfs}
\usepackage{subfigure}
\usepackage{lineno}
\usepackage{color}
\usepackage{ulem}

\shorttitle{Frequency-Resolved Lags for AGNs}
\shortauthors{Chen et al.}

\begin{document}

\title{Corona-heated Accretion-disk Reprocessing: Frequency-Resolved Lag Predictions for UV/Optical Reverberation Mapping of Active Galactic Nuclei}

\correspondingauthor{Mouyuan Sun}
\email{msun88@xmu.edu.cn}

\author[0000-0002-4765-1500]{Jie Chen}
\affiliation{Department of Astronomy, Xiamen University, Xiamen, 
Fujian 361005, China; msun88@xmu.edu.cn}
\affiliation{Department of Astronomy, School of Physics, Peking University, Beijing 100871, China}
\affiliation{Kavli Institute for Astronomy and Astrophysics, Peking University, Beijing 100871, China}
\author[0000-0002-0771-2153]{Mouyuan Sun}
\affiliation{Department of Astronomy, Xiamen University, Xiamen, 
Fujian 361005, China; msun88@xmu.edu.cn}

\author[0000-0002-2419-6875]{Zhi-Xiang Zhang}
\affiliation{Department of Astronomy, Xiamen University, Xiamen, 
Fujian 361005, China; msun88@xmu.edu.cn}

\begin{abstract}
Continuum reverberation mapping with high-cadence, long-term UV/optical monitoring of Active Galactic Nuclei (AGNs) enables us to resolve the AGN central engine sizes on different timescales. The frequency-resolved time lags of NGC 5548 (the target for the AGN STORM I campaign) are inconsistent with the X-ray reprocessing of the classical Shakura $\&$ Sunyaev disk model. Here we show that the frequency-resolved time lags in NGC 5548 can be well produced by the Corona-Heated Accretion-disk Reprocessing (CHAR) model. Moreover, we make the CHAR model predictions of the frequency-resolved time lags for Mrk 817, the source of the AGN STORM II campaign. We also obtain the frequency-resolved time lags as a function of the black-hole mass and Eddington ratio, which is valid for black-hole masses from $10^{6.5}$ to $10^9\ M_{\odot}$, and Eddington ratios from 0.01 to 1. Moreover, we demonstrate that, with the time spans of current continuum reverberation-mapping campaigns, the lag-luminosity relation of the CHAR model can be $\tau_{\mathrm{gz}}\propto L_{\mathrm{5100}}^{0.55\pm0.04}$, which is consistent with observations. Future observations can test our results and shed new light on resolving the AGN central engine. 
\end{abstract}



\section{Introduction}
 Active Galactic Nuclei (AGNs) show strong electromagnetic radiations across the entire spectrum and are powered by the accretion of a central supermassive black hole (SMBH). One popular model to describe the SMBH gas accretion in AGNs is the geometrically thin but optically thick disk \citep[i.e., the standard thin-disk model;][]{Shakura1973}. In most cases, the SMBH accretion disks are too small to be resolved spatially with current facilities. There have been many efforts to achieve high spatial resolution in the time domain. Reverberation Mapping \citep[RM;][]{Blandford1982} is a method to obtain the emission-region sizes by measuring the time lags between light curves of different wavelengths. The RM method is first widely used to measure broad-line region (BLR) sizes and estimate black-hole masses and is now widespread in determining accretion-disk sizes \citep[for a review, see][]{Cackett2021}. For the accretion-disk RM, the underlying assumption is that the flux variations at various wavelengths are driven by the reprocessing of the same variable X-ray irradiation from the corona (hereafter, the X-ray reprocessing). Hence, the observed time lag between the X-ray emission and the reprocessed one is simply the light travel time for the X-ray to reach the accretion disk. To be precise, the radiation at a given wavelength of the accretion disk is produced by numerous parts of the disk with different temperatures, and the measured time lag is, therefore, an averaged value over the entire accretion disk weighted by variability amplitudes. 

Recent intensive UV/optical continuum reverberation mappings find that some AGNs' continuum time lags are typically 2–3 times larger than the prediction from the standard thin-disk model with X-ray reprocessing \citep[e.g.,][]{Edelson2015,Fausnaugh2016,Cackett2018,Fausnaugh2018,Kara2021,Donnan2023,Kara2023}. Several models have been proposed to account for the discrepancy. Often, these models argue that the X-ray reprocessing of the standard thin disk should be revised to include the reprocessing of BLR clouds \citep{Cackett2018, Chelouche2019, Korista2019} or disk winds \citep{Sun2019}. According to the popular BLR model, the diffuse continua from the BLR clouds make a substantial or even dominant contribution to the optical variability and also continuum reverberation lags \citep[e.g.,][]{Cackett2018, Lawther2018,Sun2018b,Chelouche2019,Korista2019,Guo2022,Netzer2022}. If this is indeed the case, the observed luminosity of an AGN then does not fully characterize its accretion rate. Other works suspect that the X-ray corona has a large and highly variable scale height \citep{Kammoun2019,Kammoun2021a,Kammoun2021b,Kammoun2023}. Some models replace the X-ray reprocessing with alternative mechanisms. For instance, \cite{Gardner2017} suggest that an ``FUV torus" is formed in the inner disk and can heat the outer disk. \cite{Sun2020} propose the Corona Heated Accretion-disk Reprocessing (CHAR) model, in which the corona and disk are coupled via magnetic fields. Hence, the magnetic turbulence in the corona can change the disk heating rate and induce disk temperature fluctuations. The CHAR model can explain the larger-than-expected accretion disk sizes without X-ray reprocessing or the diffuse continuum emission from BLR \citep[see][]{Sun2020}. The anti-correlation between the ratio of the observed time lag to the standard thin-disk prediction and the luminosity favors the BLR or the CHAR model \citep{LiT2021}. Generally, the physics of the variability process in UV/optical is still an unsettled issue. 

Popular methods for measuring RM time lags are the cross-correlation function \citep[CCF;][]{Peterson1998} and the Fourier analysis \citep[for a review, see][]{Uttley2014}. The cross-correlation function measures the correlation between the light curves of two wavelengths at each lag. The Fourier analysis calculates the phase lag of the cross-power spectrum at each frequency \citep[see Section~\ref{Frequency-resolved lag} and][]{Uttley2014}. The Fourier analysis can therefore separate lags from different processes occurring on different timescales. The Fourier analysis is widely used in X-ray RMs because of the rapid variability and high cadence in X-ray light curves. The optical light curves, however, often have a low cadence and short time duration compared with the timescale of interest, making Fourier analysis challenging to apply. 

Benefiting from the high-cadence observations with \textit{Swift} \citep{Roming2005} and ground-based telescopes, and also the techniques dealing with gaps in light curves \citep[e.g., the maximum likelihood approach;][]{Miller2010,Zoghbi2013,Cackett2022}, the Fourier analysis has been applied in analyzing the AGN Space Telescope and Optical Reverberation Mapping \citep[AGN STORM I;][]{De2015} data of NGC 5548 \citep{Cackett2022}. The authors find that the UV/optical lags in NGC 5548 generally decrease with increasing frequency. This trend can not easily be described by the X-ray reprocessing of a large accretion disk alone. Instead, a standard thin disk with the diffuse continuum emission from the BLR can explain the observations. Alternatively, the CHAR model has predicted a similar trend of the frequency-resolved time lags \citep[see fig.~7 of][]{Sun2020}. Recently, the frequency-resolved UV/optical time lags of a second source, Mrk 335, were presented by \cite{Lewin2023}. Hence, we are motivated to quantitatively compare the CHAR model with the frequency-resolved time lags of NGC 5548 and Mrk 335. Mrk 817 is the target for the AGN STORM II program \citep{Kara2021}. We, therefore, aim to predict its frequency-resolved time lags, which can be tested by future AGN STORM II data. We also provide analytical functions to describe the CHAR frequency-resolved lags as a function of black-hole mass and Eddington ratio for future observational testings. 

This manuscript is formatted as follows. In Section~\ref{Frequency-resolved lag}, we introduce the frequency-resolved time lags. In Section~\ref{sect:NGC5548+Mrk817}, we present the frequency-resolved time-lag predictions of the CHAR model for three AGNs, NGC 5548, Mrk 335, and Mrk 817. In Section~\ref{sect:Trend}, we show the frequency-resolved lags as a function of black-hole mass and Eddington ratio. In Section~\ref{sect:lag-luminosity relation}, we discuss the lag-luminosity relation of the CHAR model. Summary is made in Section~\ref{sect:summary}. Note that the Schwarzschild radius $R_\mathrm{s}\equiv 2GM_\mathrm{BH}/c^2$, where $G$ and $c$ are the gravitational constant and speed of light, respectively. The Eddington luminosity is $L_{\mathrm{Edd}}=1.3\times 10^{38}\ (M_{\mathrm{BH}}/M_{\odot})\ \mathrm{erg\ s^{-1}}$.

\section{Frequency-resolved lags}
\label{Frequency-resolved lag}
Time series can be analyzed in the time and frequency domain. In reverberation mapping studies, a widely known time-domain technique used in measuring time lags between two light curves is the cross-correlation function \citep[CCF;][]{Peterson1998}. The frequency analysis technique, on the other hand, acts as a filter on light curves and estimates the time lag at each Fourier frequency or timescale \citep[see sections 2.1.2 and 2.4.3 of][]{Uttley2014}. Here, we briefly explain the mathematics of the frequency-resolved lags.

For light curves of two wavebands $x(t)$ and $y(t)$, their $n$th value of discrete Fourier transforms are as a function of the Fourier frequency $f_n=n/(N\Delta{t})$, where $N$ and $\Delta{t}$ are the numbers of data points and cadence of the light curve, respectively; i.e,
\begin{equation}
\label{eq:DFT_x}
X_n=\sum_{k=0}^{N-1}x_k \exp(-2\pi i n k/N )=A_{x,n}\exp(i\varphi _{x,n} ),
\end{equation}
\begin{equation}
\label{eq:DFT_y}
Y_n=\sum_{k=0}^{N-1}y_k \exp(-2\pi i n k/N )=A_{y,n}\exp(i\varphi _{y,n} ),
\end{equation}
where $x_k$ and $y_k$ are the $k$th values of the light curves. These discrete Fourier transforms can be expressed as an amplitude ($A_{x,n}$ or $A_{y,n}$) and a phase ($\varphi _{x,n}$ or $\varphi _{y,n}$). If $y(t)$ has an additional phase-shift $\phi_n$ with respect to $x(t)$ at frequency $f_n$, we expect that

\begin{equation}
\label{eq:Y_n}
Y_n=A_{y,n}\exp[i(\varphi _{x,n}+\phi_n)],
\end{equation}
and their cross-power spectrum
\begin{equation}
\label{eq:cpsd}
C_{xy,n}=X^*_nY_n=A_{x,n}A_{y,n}\exp(i\phi_n).
\end{equation}
Then, the time lag at frequency $f_n$ is 
\begin{equation}
T_{\mathrm{lag}}(f_n)=\phi_n/(2\pi f_n).
\end{equation} 
Using this technique, we can obtain the lags at each frequency ranging from $f_{\mathrm{min}}=1/T_{\mathrm{obs}}=1/(N\Delta{t})$ to the Nyquist frequency $f_{\mathrm{max}}=1/(2\Delta{t})$. 

Compared to CCF, the Fourier analysis provides a more straightforward way for revealing the reverberation mapping features hidden in the data. In X-ray RMs, for example, the positive time lags between soft and hard X-rays on short timescales are considered to be the light travel time between the direct power-law X-ray emission of the corona and the reflected component of the disk \citep[for a review, see][]{Uttley2014}. On the other hand, the negative time lags on long timescales indicate disk variability propagates from the disk to the corona. Strictly speaking, both the corona and disk produce soft X-ray emission, so the measured time lags on short timescales can be less than the true light travel time from the corona to the disk, and the shape of the lag-frequency relationship will be changed to some extent \citep[i.e., the dilution effect; see section 4.1.1 and fig.~20 of][]{Uttley2014}. Since the size of the accretion disk is much larger than the X-ray corona, the dilution effect in UV/optical RMs is more severe than in X-ray. Therefore, the frequency-resolved time lags of two disk continuum light curves are more complicated than in X-ray studies, which depend upon the overlapping in their emission regions and also possibly on the variability timescale (see Appendixes~\ref{appendixA} \& \ref{appendixB}). In summary, the frequency-resolved time lags can tell us the details of the physical process for variability and emission. 

To our best knowledge, the Fourier technique has been used to measure the frequency-resolved time lags of the AGN STORM I target NGC 5548 \citep{Cackett2022}, and other sources, e.g., Mrk 335 \citep{Lewin2023} and Fairall 9 \citep[but with large uncertainties;][]{Yao2023}. The frequency-resolved time lags of NGC 5548 from \cite{Cackett2022} show that the X-ray reprocessing of a large accretion disk produces excessive lags in high frequency and less lags in low frequency. Hence, it leads to attempts to use the BLR model \citep[see fig.~4 in][]{Cackett2022} or the CHAR model to account for the frequency-resolved lags. Since NGC 5548 has the most bands and highest cadences (i.e., the best-studied reverberation mapping AGN), we here choose it to test the CHAR model. We also examine the recent results of Mrk 335.

\section{Frequency-resolved lags of the CHAR model in the AGN STORM targets}
\label{sect:NGC5548+Mrk817}

\subsection{NGC 5548}
\label{sect:NGC 5548}
We calculate the frequency-resolved lags for light curves on NGC 5548 from the CHAR model of \cite{Sun2020} and compare them with the results from \cite{Cackett2022}. Here we briefly introduce the CHAR model. The CHAR model is based on the physical picture that, the chaotic magnetohydrodynamic (MHD) fluctuations in the corona propagate into the accretion disk and alter the disk MHD turbulence dissipation rate, which is also the accretion-disk gas heating rate. Then, the disk temperature fluctuates in response to the variable heating rate. The CHAR model assumes the disk keeps vertical hydrostatic equilibrium. Given the PSD of the heating-rate fluctuations and initial temperature conditions of a standard thin disk, the temperature fluctuations can be determined by solving the equation for vertical integrated thermal-energy conservation \citep[for a detailed description, see section 2 of ][]{Sun2020}. Hence, the CHAR model has three parameters introduced by the standard thin disk, namely, the black-hole mass ($M_{\mathrm{BH}}$), Eddington ratio\footnote{One of the parameters of the CHAR model is the dimensionless accretion rate $\dot{m}$. Here, $\dot{m}={\dot{M}}/{\dot{M}_{\mathrm{Edd}}}$, where the absolute accretion rate $\dot{M}=L_{\mathrm{bol}}/({\eta c^2})$, and the Eddington accretion rate $\dot{M}_{\mathrm{Edd}}=10L_{\mathrm{Edd}}/c^2$. $\eta$ is the radiative efficiency. We use $\eta=0.1$ to estimate $\dot{M}$. Hence, $\dot{m}$ also represents the Eddington ratio.} ($L_{\mathrm{bol}}/L_{\mathrm{Edd}}$, where $L_{\mathrm{bol}}$ is bolometric luminosity), and the dimensionless viscosity parameter ($\alpha$). Given these three parameters, the CHAR model can generate simulated light curves for the fixed wavelengths by integrating the multi-temperature blackbody emission over the whole accretion disk. We stress that general relativistic effects and color corrections are not considered in the calculation; hence, the results are valid for those bands whose corresponding emission regions are not close to the SMBH event horizon. Otherwise, the calculated time lags under-estimate the true time lag \citep[the demonstration of the general relativistic effects for the X-ray reprocessing is given by, e.g.,][]{Kammoun2023}, and this under-estimation on longer timescales is weaker than on short timescales. Following previous studies of NGC 5548, we fix $M_{\mathrm{BH}}=5 \times 10^7\ M_{\odot}$ \citep{Edelson2015,Fausnaugh2016} and $L_{\mathrm{bol}}/L_{\mathrm{Edd}}=0.02$ (here, we, for simplicity, assume the disk is face-on). The remaining parameter $\alpha=0.2$ \citep[e.g.,][]{King2007}. The inner and outer boundaries of the accretion disk are fixed to be $3\,R_\mathrm{s}$ and $30,000\,R_\mathrm{s}$, respectively. \textit{Hence, there are no free parameters in our calculation.} In our simulations, we adopt the redshift $z= 0.017175$ \citep{De2015} and generate the simulated light curves of all 18 UV/optical bands listed in Table 5 of \cite{Fausnaugh2016}. Each light curve spans 170 days with a cadence of 0.1 days. Our simulation is repeated 512 times to account for the statistical fluctuations due to the limited time duration. 

\begin{figure*}
\centering
\includegraphics[width=0.9\textwidth]{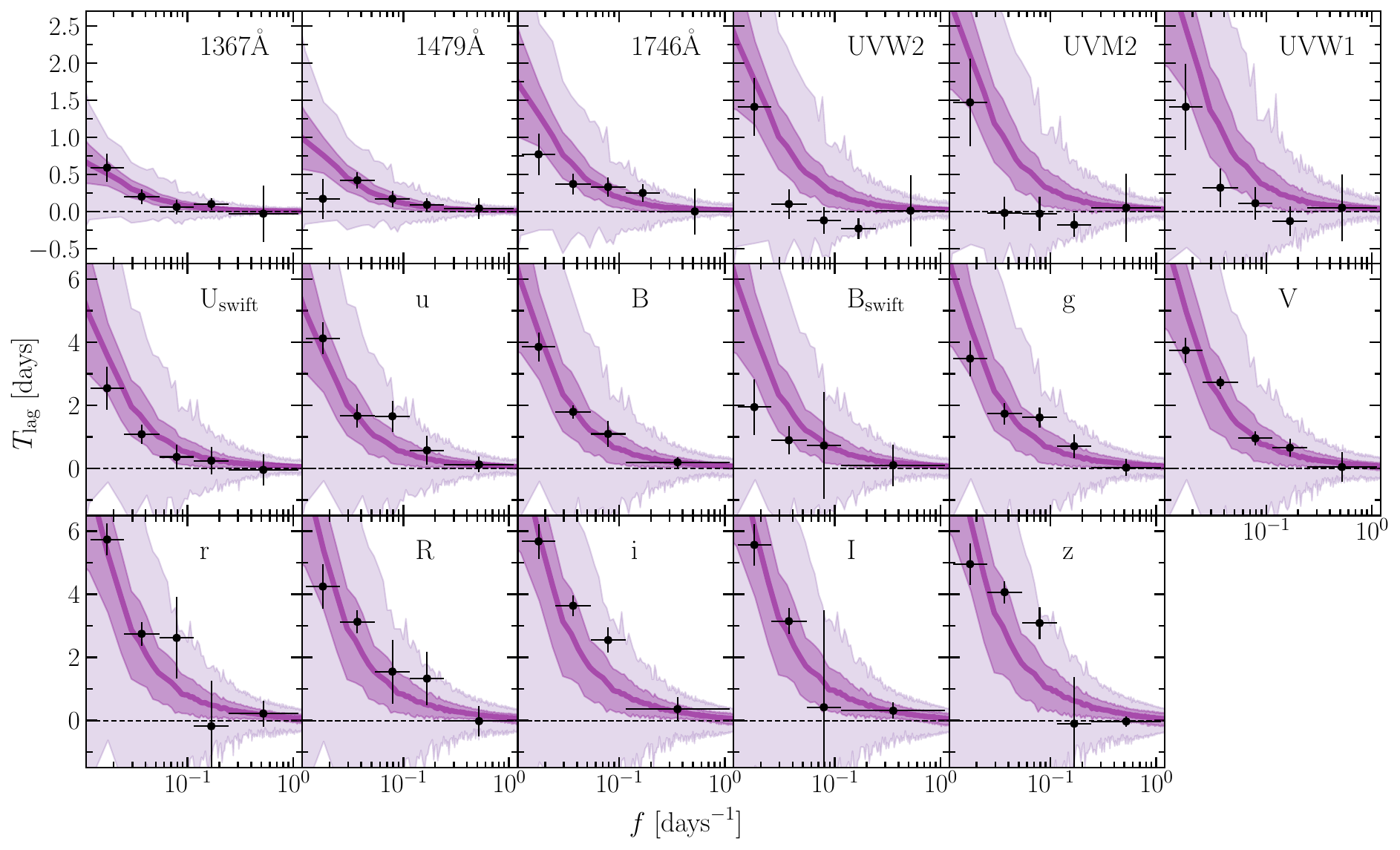}
\caption{Simulated frequency-resolved lags in NGC 5548. The frequency range is $0.011–1.21\  \mathrm{days}^{-1}$. The purple curves represent the median lags in the 512 CHAR simulations; the dark purple shaded areas correspond to the 16-th and 84-th percentiles (i.e., the $1\sigma$ uncertainties), and light purple shaded areas correspond to the 2.5-th and 97.5-th percentiles (i.e., the $2\sigma$ uncertainties). Black dots with error bars are the frequency-resolved lags and their $1\sigma$ uncertainties from \cite{Cackett2022}, who measured the time lags from the NGC 5548 AGN STORM I observations. Note that the y-axis ranges in the top panels are narrower than those of the middle and bottom panels. The simulated frequency-resolved lags for NGC 5548 from the CHAR model without free parameters are broadly consistent with observations.}
\label{fig:NGC 5548}
\end{figure*}

We adopt the Fourier technique outlined in Section~\ref{Frequency-resolved lag} to measure frequency-resolved time lags in the NGC 5548 simulated light curves. All the lags are measured with respect to the $1158\ \mathrm{\AA}$ emission \citep[i.e., the same as][]{Cackett2022} by the standard fast Fourier technique. The median lags of the 512 CHAR results are shown as the purple curves in Fig.~\ref{fig:NGC 5548}. We use the 16-th and 84-th percentiles of the 512 simulations as $1\sigma$ uncertainties (i.e., the dark purple shaded areas), and the 2.5-th and 97.5-th percentiles as the $2\sigma$ uncertainties (i.e., the light purple shaded areas).
 
Our simulated lag-frequency relation agrees well with the observations of NGC 5548 \citep{Cackett2022} for all bands in Fig.~\ref{fig:NGC 5548}. Like real observations, the simulated lags decrease with increasing frequency. Quantitatively, we use the reduced chi-square $\chi^2_{\mathrm{reduced}}$ to measure the goodness of fit between the model and observations, which is defined as the ratio of $\chi^2$ to the degrees of freedom (DOF). Here the statistic $\chi^2$ is defined as 
\begin{equation}
\label{eq:chi2}
\chi^2=\sum_{i=1}^{N}\frac{(y_{i,\mathrm{ model}}-y_{i,\mathrm{ data}})^2}{\sigma_{i,\mathrm{model}}^2+\sigma_{i,\mathrm{data}} ^2}, 
\end{equation}
where $y_{i}$ and $\sigma_{i}$ are the time lag value and its $1 \sigma$ uncertainty from real data and the model, respectively. For the CHAR model, $\chi^2 =95.01$ for 81 data points without free parameters, giving $\chi^2_{\mathrm{reduced}}=1.173$. In the modeling of the small disk plus BLR model \citep{Cackett2022}, $\chi^2 =70.95 $ for 81 data points and 17 free parameters, giving $\chi^2_{\mathrm{reduced}}=1.109$; for the X-ray reprocessing of a large accretion disk model, $\chi^2 =236.4$ for 81 data points without free parameters, giving $\chi^2_{\mathrm{reduced}}=2.919$ \citep{Cackett2022}. Hence, the CHAR model performs similarly to the small disk plus BLR model, and both models are better than the X-ray reprocessing model. 

In the CHAR model, the time lag between light curves of two wavelengths comes from two physical processes, i.e., the propagation of coronal magnetic fluctuations and the disk heating process by magnetic fluctuations. The former is near to the light travel time, and the latter is thermal timescale ($\tau_{\mathrm{TH}} \sim 1/(\alpha \Omega_{\mathrm{K}})$, where $\Omega_{\mathrm{K}}$ is Keplerian angular velocity; this is the gas temperature fluctuation timescale in the CHAR model). The thermal timescale indicates a slower temperature response for gas in the outer regions than in the inner parts, contributing large time lags on long timescales \citep{Sun2020}. Unlike the BLR model whose response function is a summation of the disk and BLR components with variable fraction, the CHAR model considers the magnetohydrodynamic of the accretion disk itself, which can explain the frequency-resolved lags of NGC 5548.

We also consider the case of $\alpha=0.01$ and find that the time lags of the model are smaller than lags from the observations, which suggests that $\alpha\simeq 0.2$ may be more consistent with real cases \citep{King2007}.

\subsection{Mrk 335}
\begin{figure*}
\plotone{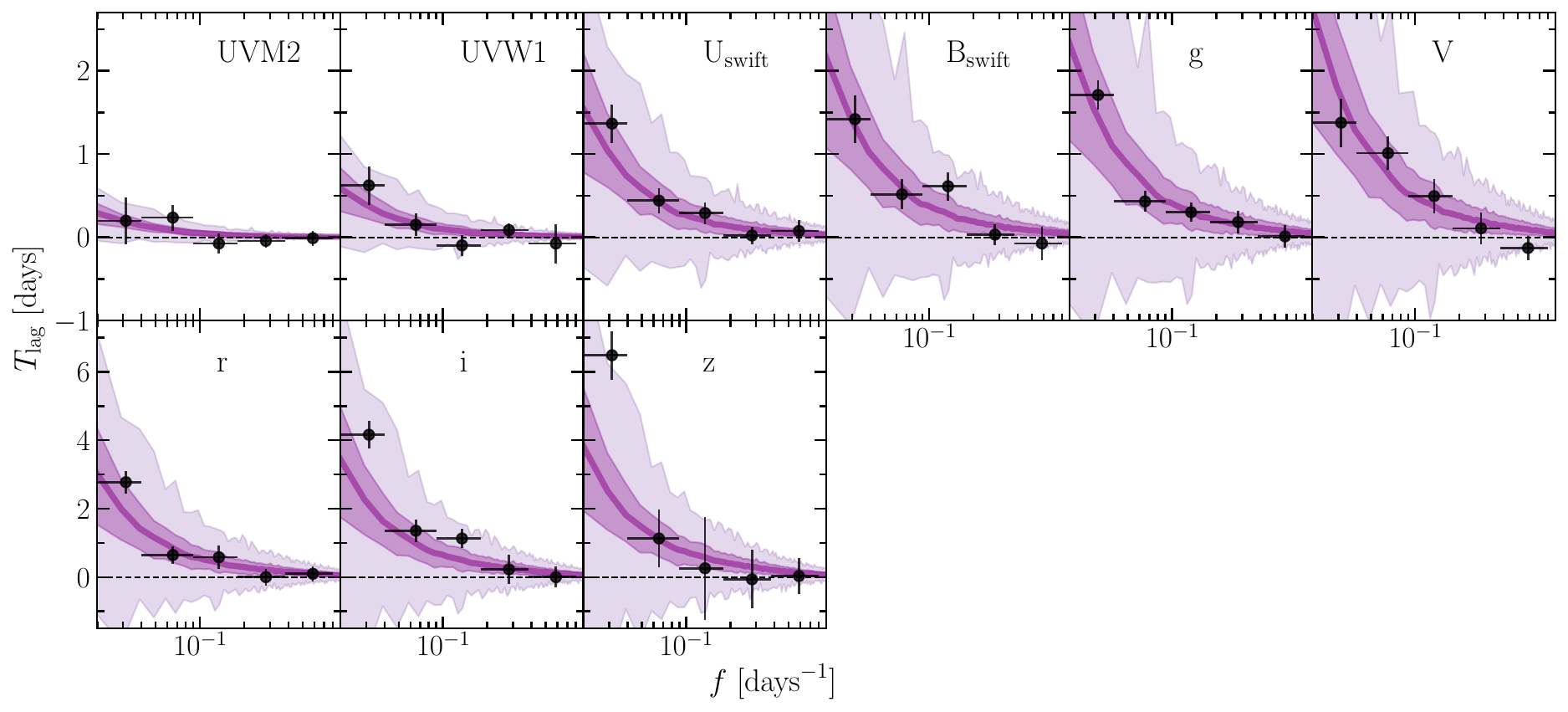}
\caption{Simulated frequency-resolved lags in Mrk 335 with respect to the UVW2 band. The frequency range is $0.02–0.9\  \mathrm{days}^{-1}$. Black dots with error bars are the frequency-resolved lags and their $1\sigma$ uncertainties from \cite{Lewin2023}. The simulated frequency-resolved lags for Mrk 335 from the CHAR model are broadly consistent with observations. }
\label{fig:mrk335}
\end{figure*}

Very recently, the frequency-resolved lags in Mrk 335 were obtained by \cite{Lewin2023}. Here we run 512 CHAR model simulations for Mrk 335 with  $M_{\mathrm {BH}} = 1.69 \times 10^7\ M_{\odot}$ \citep{Grier2012},  $L/L_{\mathrm{Edd}}= 0.07$ \citep{Tripathi2020}, and redshift $z=0.025785$ \citep{Huchra1999}. For the simulated light curves, the time duration in each band is 100 days with a cadence of 0.3 days. We present the UV/optical lag-frequency relations with respect to the UVW2 band in Fig.~\ref{fig:mrk335}.  The CHAR model agrees well with the observations. Quantitatively, $\chi^2 =49.79$ for 45 data points without free parameters, gives $\chi^2_{\mathrm{reduced}}=1.107$.

\subsection{Mrk 817}
\label{sect:Mrk 817}
\begin{figure*}
\centering
\includegraphics[width=0.9\textwidth]{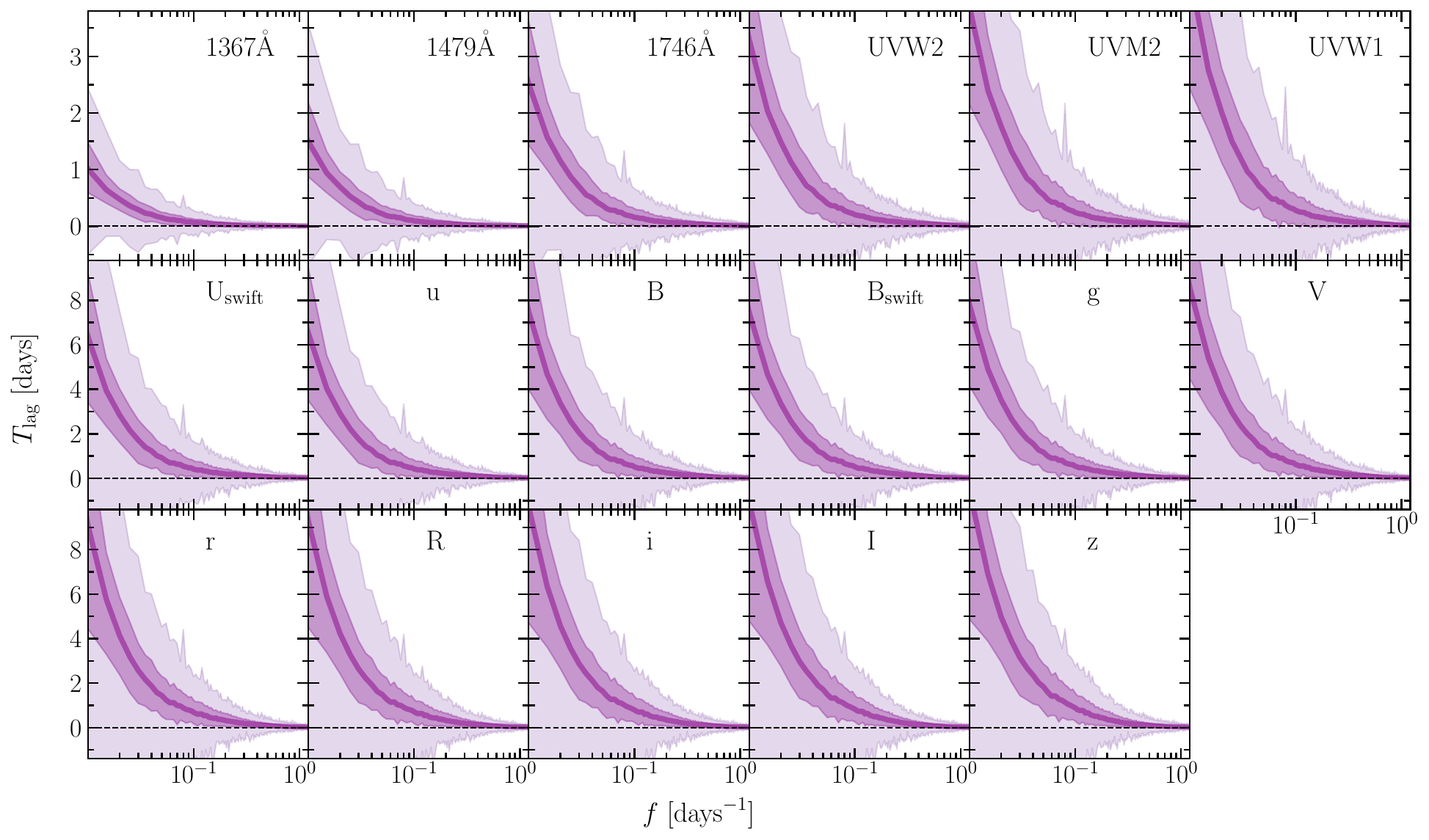}
\caption{Similar to Fig.~\ref{fig:NGC 5548}, but for Mrk 817.}
\label{fig:Mrk 817}
\end{figure*}

The target for the AGN STORM II campaign is the Seyfert 1 galaxy Mrk 817, whose mass is similar to that of NGC 5548 but more luminous. Two essential CHAR model parameters of Mrk 817 are from \cite{Kara2021}: the black-hole mass $M_{\mathrm {BH}} = 3.85 \times 10^7\ M_{\odot}$, and the Eddington ratio $L/L_{\mathrm{Edd}}= 0.2$. We run a similar simulation with the redshift $z= 0.031455$ for Mrk 817 \citep{Strauss1988}. For the simulated light curves, the time duration in each band is 200 days with a cadence of 0.1 days. We present the lag-frequency predictions for Mrk 817 in Fig.~\ref{fig:Mrk 817}, and these simulated results can be tested by the Mrk 817 AGN-STORM II observations.

\section{Trends of frequency-resolved lags in the CHAR model}
\label{sect:Trend}

In the CHAR model, the rest-frame time lag ($T_\mathrm{lag}$) is a function of the variability frequency ($f$), wavelength ($\lambda$), black-hole mass ($M_{\mathrm{BH}}$), and Eddington ratio ($\dot m$) in terms of $T_\mathrm{lag}=g(f, \lambda, M_\mathrm{BH}, \dot m)$. For giving a comprehensive prediction, we simulate 30 cases in the parameter space of $M_{\mathrm{BH}}$ and $\dot{m}$; then, we calculate the corresponding frequency-resolved lags. These 30 cases are selected as follows: $M_{\mathrm{BH}}$ ranges from $10^{6.5}$ to $10^9\ M_{\odot}$, with six values in equal logarithmic increments; $\dot{m}$ ranges from 0.01 to 1, with five values in equal logarithmic increments. We generate the simulated light curves of 18 UV/optical bands like NGC 5548. The time duration in each band is 400 days with a cadence of 0.1 days. Each case is repeated 512 times in CHAR simulations. Based on the CHAR simulated light curves of the 30 mock sources, we compute the frequency-resolved lags with respect to the $1158\ \mathrm{\AA}$ emission. The lag-frequency relations are available in the FITS file format and can be accessed from \url{https://doi.org/10.12149/101308}.

\begin{figure*}
\plotone{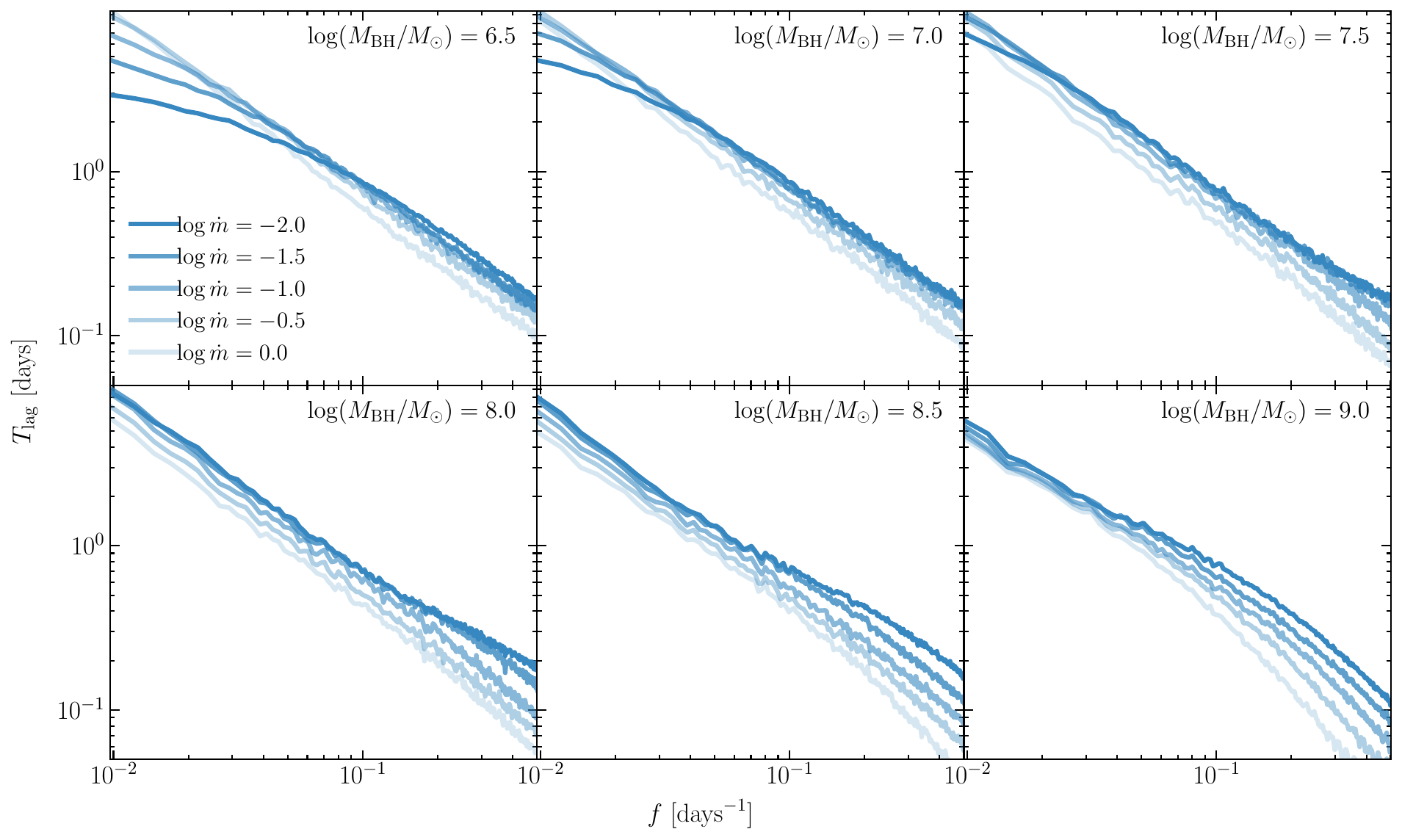}
\caption{Simulated frequency-resolved lags of 5404 $\mathrm{\AA}$ (i.e., $V$ band) vs 1158 $\mathrm{\AA}$ for the 30 cases. The frequency range is $0.097–0.5\  \mathrm{days}^{-1}$. Each panel represents the frequency-lag relation at a fixed black-hole mass for different Eddington ratios. The lighter blue color corresponds to a larger accretion rate. On the low-frequency region (e.g., $\sim 0.01\ \mathrm{days^{-1}}$), the time lag increases with increasing Eddington ratio at low mass, and this positive correlation weakens or even becomes slightly negative as mass increases.}
\label{fig:mass}
\end{figure*}

\begin{figure*}
\plotone{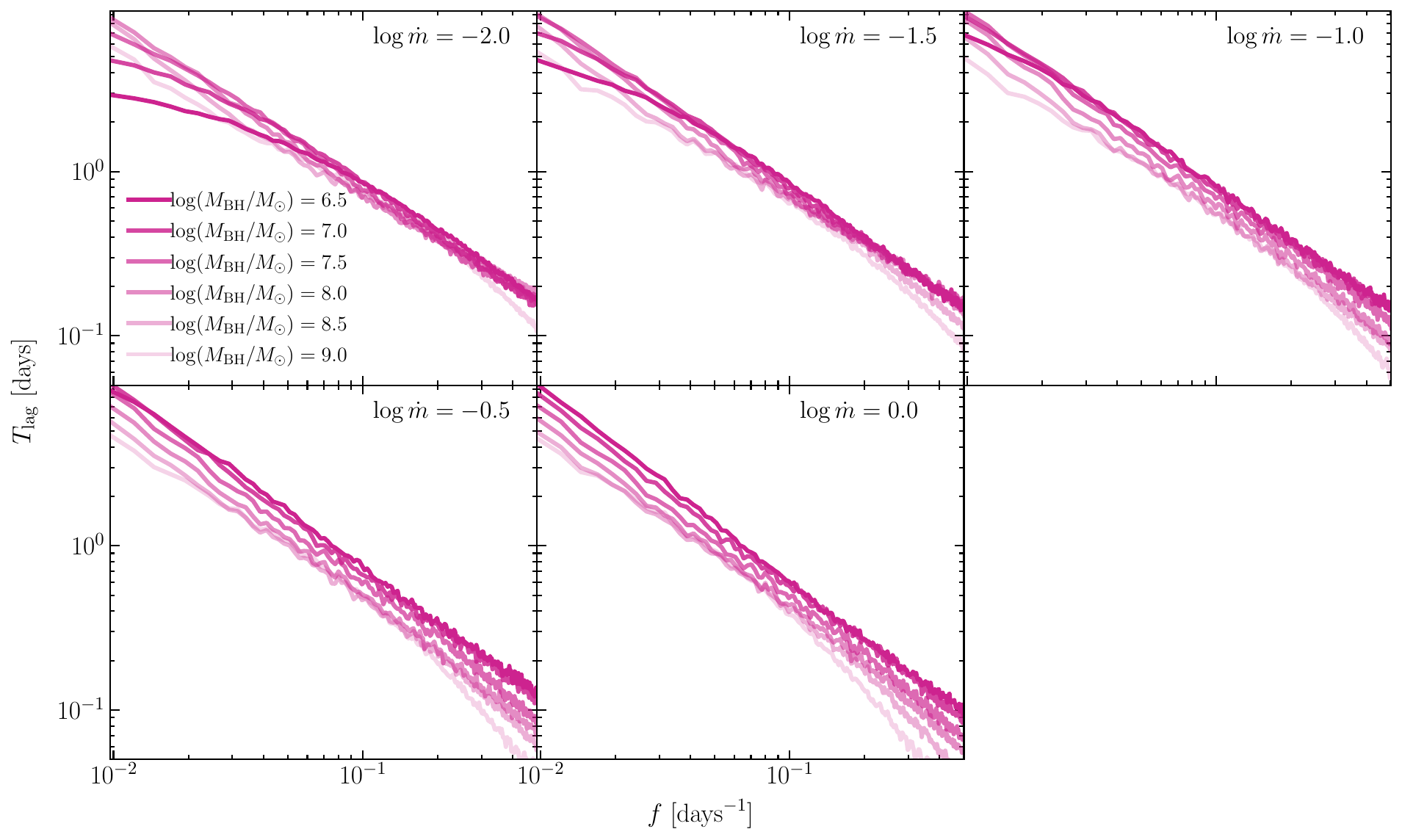}
\caption{Simulated frequency-resolved lags of 5404 $\mathrm{\AA}$ (i.e., $V$ band) vs 1158 $\mathrm{\AA}$ for the 30 cases. The frequency range is $0.097–0.5\  \mathrm{days}^{-1}$. Each panel represents the frequency-lag relation at a fixed Eddington ratio for different black-hole masses. The lighter pink color corresponds to a larger black-hole mass. On the low-frequency region (e.g., $\sim 0.01\ \mathrm{days^{-1}}$), the time lag increases with increasing mass at low Eddington ratios, and this positive correlation weakens or even becomes slightly negative as the Eddington ratio increases.}
\label{fig:ratio}
\end{figure*}

We explore the dependence of lags upon $M_{\mathrm{BH}}$, $\dot{m}$, and $f$ by considering the lags between the $V$ band $(5404\ \mathrm{\AA})$ and $1158\ \mathrm{\AA}$ emission. We show the frequency-resolved lags for the 30 mock cases in Figs.~\ref{fig:mass} and \ref{fig:ratio}. Each panel in Fig.~\ref{fig:mass} shows the frequency-lag relation at a fixed mass for different Eddington ratios; each panel in Fig.~\ref{fig:ratio} illustrates the frequency-lag relation at a fixed Eddington ratio for different black-hole masses. For clarity, the error bars are not shown.

\begin{figure}
\plotone{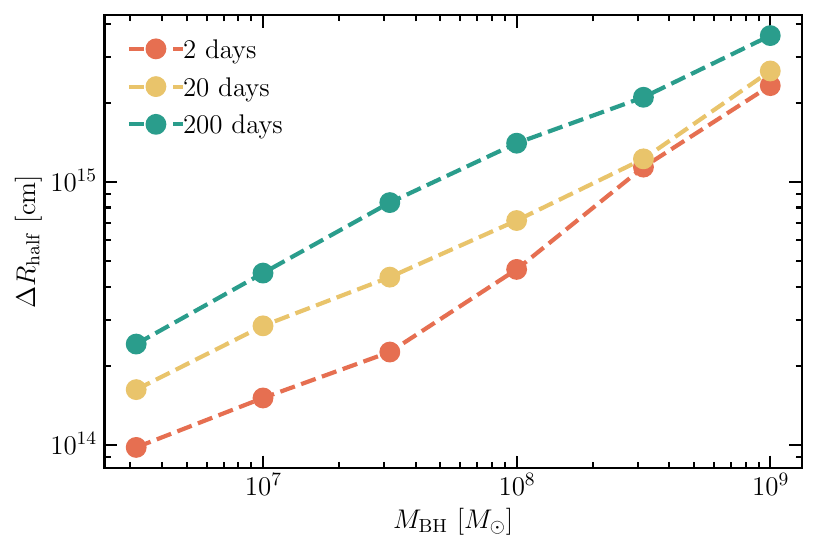}
\caption{The difference between $R_\mathrm{half,5404}$ and $R_\mathrm{half,1158}$ in a range of black-hole masses with $\dot m=0.1$. The timescale is the time duration of the variable flux. The longer timescale and larger black-hole mass result in larger $\Delta R_\mathrm{half}$. }
\label{fig:Delta_R_50}
\end{figure}

\begin{figure}
\plotone{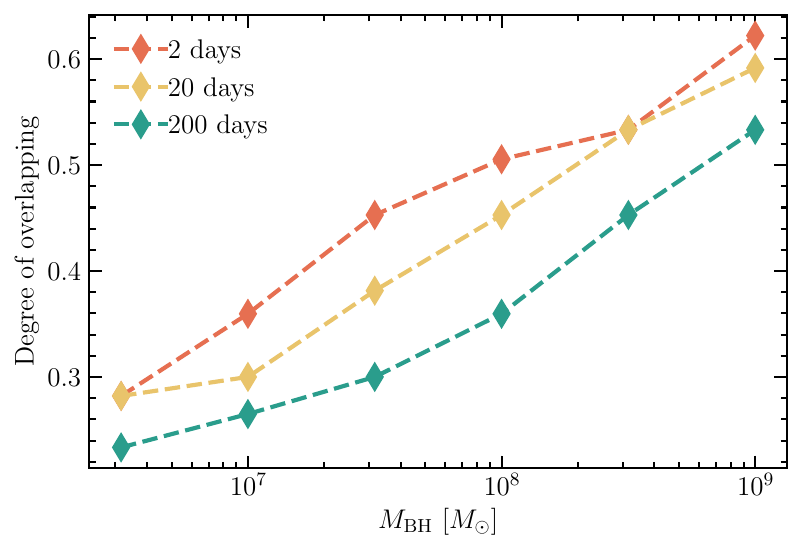}
\caption{Degree of the emission-region overlapping between the 5404 $\mathrm{\AA}$ (i.e., $V$ band) and 1158 $\mathrm{\AA}$ emission in a range of black-hole masses with $\dot m=0.1$. The shorter timescale and larger black-hole mass result in more overlapping.}
\label{fig:emission_ratio}
\end{figure}

On long timescales or low frequency (e.g., $\sim 0.01\ \mathrm{days^{-1}}$), the relation between the time lags and Eddington ratio or luminosity, as the $M_{\mathrm{BH}}$ increases, changes from positive to nearly negative. We discuss this interesting phenomenon, and the possible reasons are as follows. Qualitatively, as black-hole mass or Eddington ratio increases, the emission regions of a given wavelength increase due to the larger accretion disk or higher effective temperature, respectively. To quantify the dominant position of the emission regions for a given wavelength, we calculate the half-light radius ($R_\mathrm{half}$) of the time-variable emission following \cite{Tie2018}. First, we obtain the first-order Taylor expansion of the Planck function at each radius \citep[i.e., $\Delta B(\lambda,R)$; eq.~19 from][]{Sun2020}, which represents the variation of intensity as a function of radius. We then calculate the cumulative contribution fraction of $\leq R$ regions to the total variability by
\begin{equation}
    f_{\Delta \mathrm{L}}(\lambda, R)= \frac{\int_{ R_{\mathrm{in}}}^{R} \Delta B\left(\lambda, r\right) r \mathrm{d} r}{\int_{ R_{\mathrm{in}}}^{R_{\mathrm{out}}} \Delta B\left(\lambda, r\right) r \mathrm{d} r}\\,
\end{equation}
where $R_{\mathrm{in}}$ and $R_{\mathrm{out}}$ are the inner and outer boundaries of the accretion disk, respectively. The half-light radius of the variable flux is calculated by setting $f_{\Delta \mathrm{L}}(\lambda, R_\mathrm{half})= 0.5$. We obtain $\Delta R_\mathrm{half}=R_\mathrm{half,5404}-R_\mathrm{half,1158}$ in a range of black-hole masses with $\dot m=0.1$ in Fig.~\ref{fig:Delta_R_50}. Consistent with our expectations, the relative distance between the emission regions of the two wavelengths increases with black-hole mass.

The time lag, however, does not necessarily increase with the relative distance. The first reason is the dilution effect, which is caused by the fact the emission regions of the shorter wavelength can significantly overlap with those of the longer wavelength \citep{Uttley2014}. Hence, the measured lag of the two bands depends upon not only the relative distance but also the overlapping of their emission regions. The larger overlapping gives a shorter lag; an example is presented in Appendix~\ref{appendixA}. Here we define the overlapping degree as 
\begin{equation}
\label{eq:overlap}
p_{\mathrm{ov}}=1-\frac{R_\mathrm{half,5404}-R_\mathrm{half,1158}}{R_\mathrm{half,5404}+R_\mathrm{half,1158}}.
\end{equation}
We can obtain the overlapping degrees on different timescales. The relationship between overlapping and $M_{\mathrm{BH}}$ is shown in Fig.~\ref{fig:emission_ratio}. It shows that the shorter timescale and larger $M_{\mathrm{BH}}$ result in a higher degree of overlapping and, therefore, smaller lags in high-luminosity sources. The frequency-resolve time lags also depend upon the variability timescale. We find that the larger variability timescale also gives shorter measured lags (for a detailed discussion, see Appendix~\ref{appendixB}). Since the thermal timescale at a fixed radius increases with mass or luminosity, the times lags in high-luminosity sources will decrease accordingly. In summary, these two effects can explain the smaller lags in large black-hole mass or high-luminosity sources.

To find the mathematical form of $T_\mathrm{lag}=g(f, \lambda, M_\mathrm{BH}, \dot m)$ from our CHAR simulations, we assume the function takes the following form for each $M_{\mathrm{BH}}$ and $\dot{m}$,
\begin{equation}
T_\mathrm{lag}=(\frac{\lambda}{\lambda_0}-1)^\beta \frac{T_0}{f+f_0},
\label{eq:tlag}
\end{equation}
where $\lambda_0=1158\ \mathrm{\AA}$. The coefficients $\beta$, $f_0$, and $T_0$ depend upon $M_{\mathrm{BH}}$ and $\dot{m}$. In our fitting, $T_\mathrm{lag}$, $f$, and $M_{\mathrm{BH}}$ are in units of days, $\mathrm{days}^{-1}$, and $ M_{\odot}$, respectively. The frequency range in our fitting is $0.0073–4.47\  \mathrm{days}^{-1}$. At very high frequencies (e.g., $f>2\ \mathrm{days^{-1}}$ for $M_\mathrm{BH}=10^{6.5}\  M_{\odot}$; $f>1\ \mathrm{days^{-1}}$ for $M_\mathrm{BH}=10^{7}\ M_{\odot}$), the correlation between $T_\mathrm{lag}$ and $\lambda$ is weak. To reject these frequencies from the fitting, we calculate the Pearson correlation coefficient ($r$) between $T_\mathrm{lag}$ and $\log (\lambda/\lambda_0-1)$ for each frequency, $M_\mathrm{BH}$, and $\dot m$. For each combination of $M_{\mathrm{BH}}$ and $\dot{m}$, we select frequency-resolved lags with $r^2$ greater than 0.9 and then fit them with Eq.~\ref{eq:tlag}. We obtain the best-fitting $\beta,~ f_0,$ and $T_0$ for each $M_ {\mathrm {BH}}$ and $\dot m $ combination. We then use the second-order polynomials to fit the relations among these coefficients, $\log (m)\equiv \log(M_{\mathrm{BH}}/M_{\odot})$, and $\log(\dot{m})$. The best-fitting results are
\begin{eqnarray}
\beta=-0.2792\log (m)-0.0248\log (\dot{m})\nonumber \\ +0.0161\log^2 (m)+0.0156\log^2 (\dot{m})+1.7438,
\end{eqnarray} 
\begin{eqnarray}
f_0=-0.0441\log (m)+0.0014\log( \dot{m})\nonumber \\ 
+0.0029\log^2 (m) +0.0025\log^2 (\dot{m})+0.1637,
\end{eqnarray} 
\begin{eqnarray}
T_0=0.0038\log (\dot{m})+0.0011\log^2 (\dot{m})+0.0186.
\end{eqnarray} 
These can be used for predicting the CHAR model lags of sources with various $M_{\mathrm{BH}}$ and $\dot m$. 

We also use the supervised machine learning algorithm, Random Forest \citep[hereafter RF;][]{Breiman2001}, to learn the relation $T_{\mathrm{lag}} = g(f, \lambda, M_{\mathrm{BH}}, \dot{m})$. In the Random Forest, a decision tree is a hierarchical structure that makes successive partitions of the data with a set of if-then-else decision rules. The random forest algorithm is an ensemble technique that combines multiple decision trees on various sub-samples of the data set and uses averaging to improve the predictive accuracy and control over-fitting.

We used the scikit-learn \citep{Pedregosa2011} Python package implementation of RF\footnote{\url{https://scikit-learn.org/stable/modules/generated/sklearn.ensemble.RandomForestRegressor} }, building multiple regression model on $T_\mathrm{lag}$ with $f,\lambda, M_\mathrm{BH}$, and $\dot m $ as independent variables in logarithmic space. Again, we only consider frequencies whose corresponding time lags correlate with $\lambda$ (i.e., $r^2>0.9$). We then randomly split the simulated lags into training and testing sets of 70\% and 30\%, respectively. We present our best hyperparameters in Table~\ref{t_hp} obtained by \texttt{sklearn.model selection.GridSearchCV} (parameters that are not mentioned in the table are left as default values). We use the $R^2$ score and the mean squared error (MSE) to evaluate the predictions of the model. The $R^2$ and MSE of the training set and testing set are in Table~\ref{t_r}. We also use NGC 5548 and Mrk 817 to test machine learning prediction. Overall, the machine learning result is good and accurate, and they can also be used to predict the frequency-resolved lags in various AGNs. The machine-learning model is slightly more accurate than the analytical results. Our machine-learning model is packaged as a PKL file and can be downloaded from \url{https://doi.org/10.12149/101308}. 

\begin{deluxetable}{ccccc}
\tablecaption{Hyperparameters setting for RF\label{t_hp}}
\tablehead{\colhead{Model} & \colhead{Hyperparameters} & \colhead{Value}  \\ } 
\startdata
RF & n\_estimators &  80\\
& oob\_score & \emph{True} \\
\enddata
\tablecomments{Other parameters not mentioned
are left as default.}
\end{deluxetable}

\begin{deluxetable}{ccccc}
\tablecaption{Evaluation of RF\label{t_r}}
\tablehead{\colhead{Set} & \colhead{$R^2$} & \colhead{MSE}  \\ } 
\startdata
Training set &  0.9991 &  0.0003\\
Testing set &  0.9938 &  0.0018\\
NGC 5548 & 0.998 & 0.001\\	
Mrk 817 & 0.996 &0.002\\
\enddata
\tablecomments{$R^2$ and MSE are close to 1 and 0, respectively, indicating the RF model is accurate.}
\end{deluxetable}

\section{Continuum lag-luminosity relation}
\label{sect:lag-luminosity relation}

\begin{figure*}
\centering
\subfigure[]{
\includegraphics[scale=0.6]{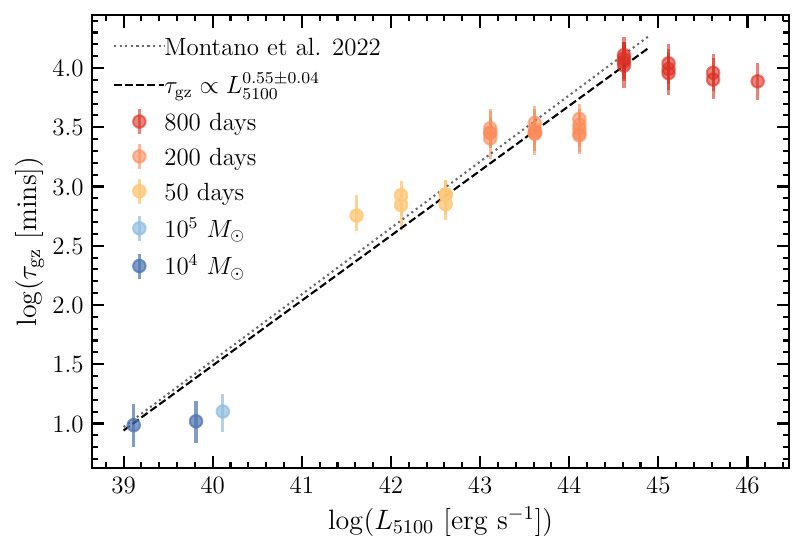}
\label{fig:L5100(a)}}
\subfigure[]{
\includegraphics[scale=0.6]{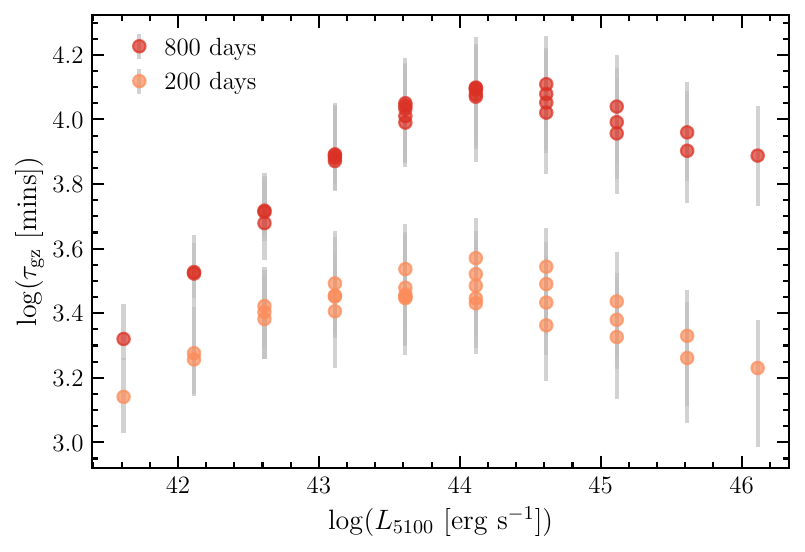}
\label{fig:L5100(b)}}
\caption{(a) $\tau_{\mathrm{gz}}$ vs $L_{5100}$ from the CHAR model for the 30 mock cases mentioned in Section~\ref{sect:Trend} and 3 additional low-mass cases introduced in Section~\ref{sect:lag-luminosity relation}. Their $M_{\mathrm{BH}}$ and $\dot{m}$ are mentioned in the first paragraph of Section~\ref{sect:Trend} and the second paragraph of Section~\ref{sect:lag-luminosity relation}, respectively. The error bars are 25-th and 75-th percentiles. The black dashed line is the best-fitting to data points with $10^{39}< L_{5100}<10^{45} \ \mathrm{erg\ s^{-1}}$. The grey dotted line is the observational fitting result from \cite{Montano2022}. Note that several data points overlap at the same luminosity and have different $M_\mathrm{BH}$ and $\dot m$. (b) $\tau_{\mathrm{gz}}$ vs $L_{5100}$ for time spans of 200 days and 800 days for the 30 mock cases mentioned in Section~\ref{sect:Trend}. A longer time span results in a larger time lag; a decreasing trend appears at high luminosity in the same time span; the critical luminosity, below which this decreasing trend shows, is associated with the time span.}
\label{fig:L5100}
\end{figure*}

The continuum time lag-luminosity relation can probe the origin of AGN UV/optical variability. It is often argued that the observed continuum time lag-luminosity relation with a slope of 0.5 resembles the classical BLR radius-luminosity relation \citep{Bentz2013}. Hence, it is proposed that the observed continuum lags are echoes of the diffuse continua from inner BLR clouds \citep[e.g.,][]{ Guo2022,Montano2022,Netzer2022}. In actual UV/optical observation, the luminous quasars are observed for longer time spans (i.e., the length of time for the monitoring campaign) than their faint counterparts because the formers are generally less variable than the latters \citep[e.g.,][]{MacLeod2010, Sun2018b}. For example, for the low-luminosity source, NGC 4593 ($L_\mathrm{bol}=7.8\times10^{43}\ \mathrm{erg\ s^{-1}}$), the time span of the time-lag observations is 22.6 days \citep{McHardy2018} in the rest-frame; for NGC 5548 ($L_\mathrm{bol}=3.2\times10^{44}\ \mathrm{erg\ s^{-1}}$), the time span is about 200 days \citep{Edelson2015, Fausnaugh2016} in the rest-frame; for luminous quasars \citep[e.g., ][]{Guo2022}, the time span (in rest-frame) is about 800 days in the rest-frame. Inspired by the anti-correlation between the variability frequency and time lag, the differences in the time spans can induce the following observational bias: in order to measure inter-band time lags, the reverberation-mapping campaigns of luminous sources require longer time spans and thus have longer time lags than faint ones. We speculate that this bias may be responsible for the observed lag-luminosity relation. Hence, we are motivated to check the time lag-luminosity relation of the CHAR model.

We use the 30 mock cases, whose $M_{\mathrm{BH}}$ and $\dot{m}$ are introduced in the first paragraph of Section~\ref{sect:Trend}, to calculate the time lag-luminosity relation of the CHAR model. We obtain the continuum luminosity $L_{5100}=\lambda L_{\lambda}$ at $5100\ \mathrm{\AA}$, which is empirically converted by $L_{\mathrm{bol}}=10L_{5100}$ \citep{Hopkins2007}. We split 30 mock sources into three luminosity bins and set the time span of 50 days in the low luminosity ($L_{5100}<10^{43}\ \mathrm{erg\ s^{-1}}$), 200 days in the middle luminosity ($10^{43}<L_{5100}<5\times10^{44}\ \mathrm{erg\ s^{-1}}$), and 800 days in the high luminosity ($L_{5100}>5\times10^{44}\ \mathrm{erg\ s^{-1}}$). This setting is roughly consistent with the time spans of real continuum reverberation mapping programs in these luminosity ranges. The cadence of simulated light curves is 0.5 days. To extend the low-luminosity end to cover NGC 4395 \citep[$L_\mathrm{bol}=5.3\times10^{40}\ \mathrm{erg\ s^{-1}}$;][]{Moran2005}, we consider low-mass mock AGNs of $M_{\mathrm{BH}}=10^4\ M_{\odot}$, $\dot{m}=0.01$ and 0.05, and $M_{\mathrm{BH}}=10^5\ M_{\odot}$, $\dot{m}=0.01$. The time span for the three low-mass mock AGNs is one day \citep[e.g.,][]{Montano2022} with a cadence of 0.005 days (i.e., seven minutes). Note that all mentioned timescales are in the rest frame. We use the CHAR model to generate multi-band light curves for these mock sources.

To compare the lag-luminosity relation of the CHAR model with observations \citep[e.g.,][]{Guo2022, Montano2022,Netzer2022}, we use \texttt{PYCCF} \citep{PYCCF} to measure CHAR model mock light curves. Following \cite{Montano2022}, we calculate the inter-band lag ($\tau$) of the $z$ band (9157\ $\mathrm{\AA}$) with respect to the $g$ band (4476\ $\mathrm{\AA}$). We always perform a first-order polynomial detrending for all light curves prior to lag measurements to remove the long-term trend \citep{Welsh1999}.

We obtain the median model lags and their $1\sigma$ uncertainties from the $128$ CHAR simulations, as shown in Fig.~\ref{fig:L5100(a)}. In the range of $10^{39}< L_{5100}<10^{45} \ \mathrm{erg\ s^{-1}}$, we use the least square method to fit a linear relationship between time lags and luminosity in logarithmic space, 
\begin{equation}
\label{eq:tlag-L5100}
\log( \tau_{\mathrm{gz}}/\mathrm{mins})= a \log( L_{5100}/\mathrm{erg~s^{-1}})+b.
\end{equation}
The best-fitting results and their $1\sigma$ uncertainties are $a=0.55^{+0.04}_{-0.04}$, $b=-20.44^{+1.42}_{-1.42}$. The uncertainties of the slope and intercept are calculated following \cite{Hogg2010}, assuming that the distributions of measured time lags are normal. The slope and intercept are almost identical to the observational results for local AGNs from \cite{Montano2022}, who obtained $a=0.56^{+0.05}_{-0.04}$, $b=-20.87^{+1.88}_{-1.96}$. Hence, the CHAR model can reproduce the lag-luminosity relation with a slope of $\sim 0.5$ without the diffuse continua from the BLR.

To discuss the relationship between the CCF lags and time span, we set the same time span of 30 mock sources mentioned in Section~\ref{sect:Trend} to 200 days and 800 days, respectively. We calculate the corresponding time lag $\tau_{\mathrm{gz}}$. The median model lags and their $1\sigma$ uncertainties from the 128 CHAR simulations are presented in Fig.~\ref{fig:L5100(b)}. It shows that the time span significantly impacts the lag-luminosity relation, i.e., a longer time span results in a larger time lag. This is because the signal with a longer time span contains more low-frequency components. Fig.~\ref{fig:L5100(b)} also shows that the lag-luminosity relation changes if the same time span is set. Overall, the tests in Fig.~\ref{fig:L5100} indicate that the continuum lag-luminosity relation from observations may be partly caused by the differences in the time spans, i.e., the luminous sources are designed to have longer time-span observations than those of faint ones because quasar variability anti-correlates with luminosity \citep[e.g.,][]{Kelly2009,MacLeod2010,Sun2018c,Suberlak2021}.

Another striking feature in Fig.~\ref{fig:L5100(b)} is the decreasing trend in the time lags for high-luminosity mock sources ($L_{5100}>10^{44}\ \mathrm{erg\ s^{-1}}$) under the same time span; this tendency deviates from the positive correlation for less luminous sources. The possible reasons for decreasing the time lags are the increasing overlap of the emission regions due to the larger $M_\mathrm{BH}$, and the larger variability timescale in luminous sources (see Appendix~\ref{appendixA} \& \ref{appendixB}). The critical luminosity from the positive to the negative trend depends upon the time span. For example, it occurs near $L_{5100}\simeq 10^{44}\ \mathrm{erg\ s^{-1}}$ in 200 days, and near $L_{5100}\simeq 4\times10^{44}\ \mathrm{erg\ s^{-1}}$ in 800 days. 

Within the existing AGN observations, there are limited time-lag measurements for high-luminosity sources with $L_{5100}$ greater than $10^{45}\ \mathrm{erg\ s^{-1}}$. Hence, this decreasing trend has not yet been critically tested. For luminous AGNs, the time-lag measurement is more difficult than faint targets because the variations decrease with increasing luminosity \citep[e.g.,][]{Kelly2009,MacLeod2010,Sun2018c,Suberlak2021}. In addition, luminous AGNs are often at high redshift, and one should use infrared light curves to probe the rest-frame $g$ and $z$ bands. In general, the time-lag measurements of high-luminosity sources should be considered in future observations like LSST \citep{Ivezi2019}. Unlike the CHAR model, the BLR diffuse continuum model does not have such a critical luminosity. Hence, future LSST observations can be used to distinguish the two models.

\section{Summary}
\label{sect:summary}
We have made the CHAR model predictions for the frequency-resolved lags in UV/optical reverberation mappings. 
Our main results can be summarized as follows. 
\begin{enumerate}
\item We have used the simulated light curves of the CHAR model to reproduce the frequency-resolved lags of NGC 5548 and Mrk 335 quantitatively (see Figures~\ref{fig:NGC 5548} and \ref{fig:mrk335}; Section~\ref{sect:NGC5548+Mrk817}).

\item We have made predictions for the frequency-resolved lags in Mrk 817 (the target for the AGN STORM II program) and for other sources with a range of black-hole masses and Eddington ratios. We have obtained the time lags as a function of the variability frequency, wavelength, black-hole mass, and Eddington ratio (see Figures~\ref{fig:Mrk 817}, \ref{fig:mass} and \ref{fig:ratio}; Sections~\ref{sect:NGC5548+Mrk817} and \ref{sect:Trend}).
 
\item The continuum lag-luminosity relation obtained from observations may be partly caused by the differences in the time span, i.e., the reverberation-mapping campaigns of luminous sources are designed to have longer time spans than those of faint ones because quasar variability anti-correlates with luminosity. Hence, the continuum time lag-luminosity relation might have nothing to do with the BLR (see Figure~\ref{fig:L5100}; Section~\ref{sect:lag-luminosity relation}). 
 
\item For the same time span, the time lags are positively correlated with luminosity and show a decreasing trend at high luminosity. The critical luminosity in the decreasing trend increases with the time span (see Figure~\ref{fig:L5100(b)}; Section~\ref{sect:lag-luminosity relation}).

\end{enumerate}

Therefore, the CHAR model can explain many observational facts about AGN UV/optical reverberation mappings, i.e., the frequency-resolved lags for NGC 5548 and Mrk 335. We find that the measured lags in UV/optical reverberation mappings are influenced by many factors, such as observation duration (frequency), variability timescale, and overlapping of emission regions. Therefore, it may be inaccurate to use time lags to simply represent the size of the accretion disk, especially for high-luminosity sources. The continuum lag-luminosity relation from current observations needs to be tested more rigorously. More future observations can help us critically test the CHAR model and understand AGN physics.

\section*{Acknowledgements}
We thank Shuo Liu for the helpful discussion about machine learning. We thank the referee for his/her useful comments that improve the manuscript. C.J. and M.Y.S. acknowledge support from the National Natural Science Foundation of China (NSFC-11973002), the Natural Science Foundation of Fujian Province of China (No. 2022J06002), and the China Manned Space Project grant (No. CMS-CSST-2021-A06 and CMS-CSST-2021-B11). Z.X.Z. acknowledges support from the National Natural Science Foundation of China (NSFC-12033006 and NSFC-12103041). C.J. acknowledges the support from the Undergraduate Innovation Program of Xiamen University.
\begin{acknowledgements}

\software{AstroML \citep{VanderPlas2012}, Jupyter notebooks \citep{Perez2007}, Matplotlib \citep{Hunter2007}, Numpy \& Scipy \citep{scipy}, PYCCF \citep{PYCCF}, SCIKIT-LEARN \citep{Pedregosa2011}.}
\end{acknowledgements}

\clearpage
\appendix
\section{The overlapping effect in the measured lags}
\label{appendixA}
\begin{figure}
\centering
\subfigure[]{
\includegraphics[scale=0.6]{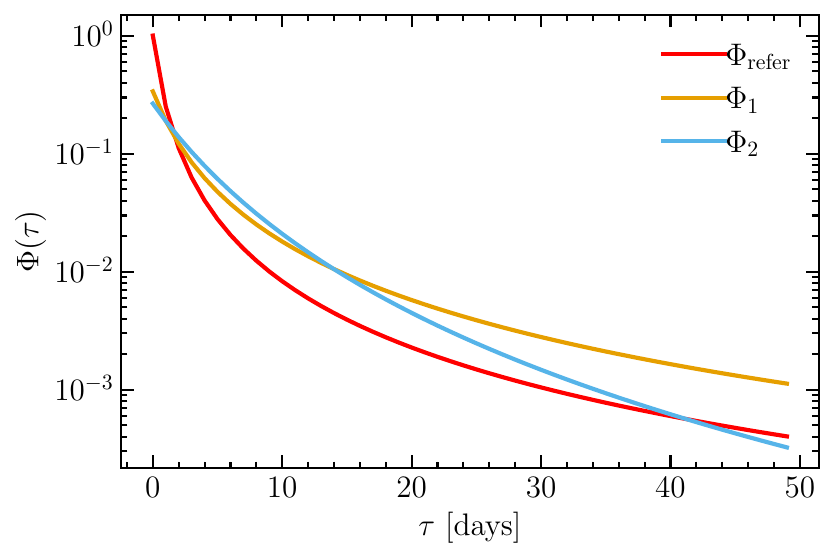}
\label{fig:DRW13(a)}}
\subfigure[]{
\includegraphics[scale=0.6]{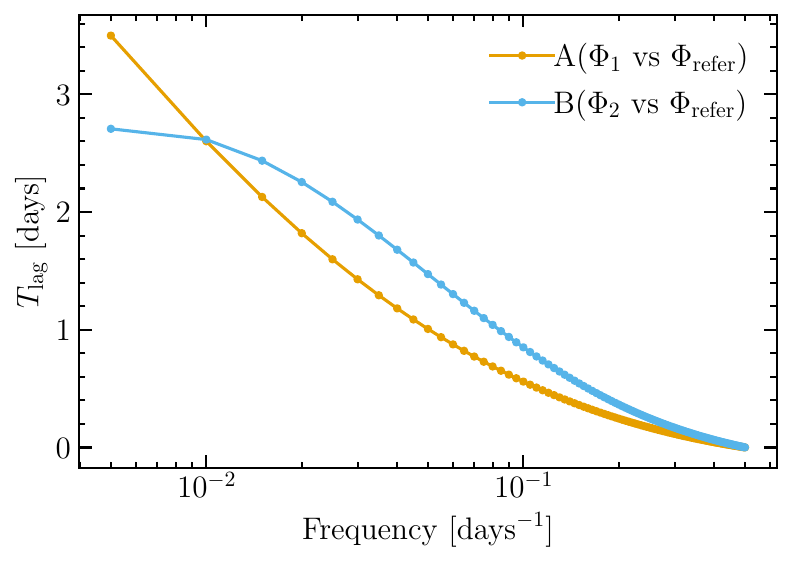}
\label{fig:DRW12(b)}}
\caption{(a) Three types of response functions. Three response functions have centroids of 4.33, 9.83, and 5.58 days and medians of 0.99, 2.91, and 2.92 days, respectively. (b) The analytical frequency-resolved lags in two cases, A($\Phi_\mathrm{1}$ vs $\Phi_\mathrm{refer}$) and B($\Phi_\mathrm{2}$ vs $\Phi_\mathrm{refer}$). At frequencies higher than $1/(2\pi \times 14)\ \mathrm{days^{-1}}$, the time lags of case A are lower than case B because the response function $\Phi_1$ is less extended than $\Phi_2$. }
\label{fig:DRW123}
\end{figure}

In a linearized RM model, the observed light curves are the convolution of the same underlying variable signal with the response function $\Phi(\tau)$ \citep{Blandford1982}. Given the response functions of two light curves, we can calculate the frequency-resolved lags by eq.~10 from \cite{Cackett2022}, i.e., the lags of the product of the Fourier transforms of two response functions. To show the diluted effect in the measured lags, we consider three types of normalized response functions shown in Fig.~\ref{fig:DRW13(a)}, $\Phi_\mathrm{refer}(\tau)=1.005/(\tau+1)^2$, $\Phi_\mathrm{1}(\tau)=3.045/(\tau+3)^2$, and $\Phi_\mathrm{2}(\tau)=4272.13/(\tau+11.25)^4$, respectively. The range for $\tau$ is from 0 to 200 days, and the constants on the numerators are the corresponding normalized coefficients. Three response functions have centroids of 4.33, 9.83, and 5.58 days, and medians (i.e., the cumulative contribution fraction equals 0.5) of 0.99, 2.91, and 2.92 days, respectively. $\Phi_\mathrm{refer}$ represents the response function of the short wavelength, while $\Phi_\mathrm{1}$ and $\Phi_\mathrm{2}$ are for long wavelength but have different distributions. $\Phi_\mathrm{1}$ have a larger centroid lag than $\Phi_\mathrm{2}$. For $\tau<14~\mathrm{days}$, $\Phi_\mathrm{2}$ is more extended than $\Phi_1$, while for $\tau>14~\mathrm{days}$, $\Phi_1$ is more extended than $\Phi_2$. We calculate the frequency-resolved lags between $\Phi_\mathrm{refer}$ and $\Phi_\mathrm{1}$ (hereafter case A) or between $\Phi_\mathrm{refer}$ and $\Phi_\mathrm{2}$ (hereafter case B), shown in Fig.~\ref{fig:DRW12(b)}. At frequencies greater than $1/(2\pi \times 14)\ \mathrm{days^{-1}}$, the lags in case A are systematically smaller than in case B. At frequencies smaller than $1/(2\pi \times 14)\ \mathrm{days^{-1}}$, the lags in case A are larger than in case B. Overall, if the emission regions of the short wavelength significantly overlap with those of the long wavelength, the frequency-resolved lags will be reduced over a wide frequency range.

\section{The variability timescale affects the measured lags}
\label{appendixB}
\begin{figure}
\centering
\subfigure[]{
\includegraphics[scale=0.6]{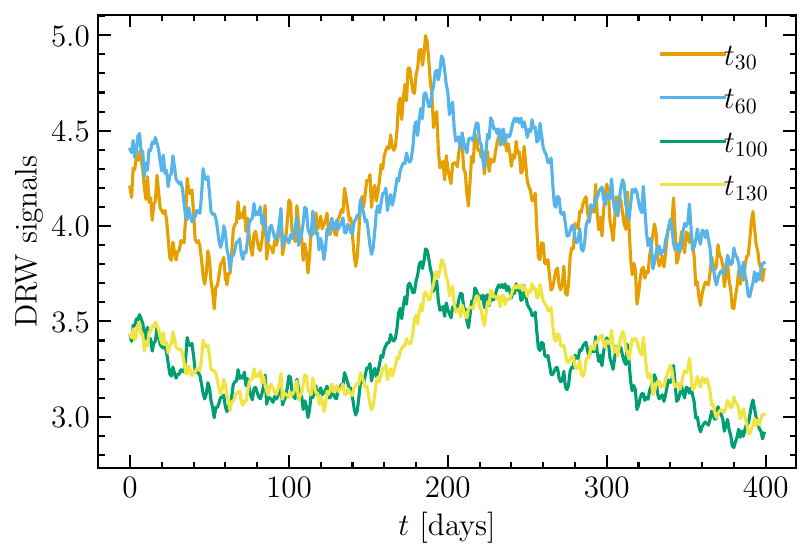}
\label{fig:DRW101(a)}}
\subfigure[]{
\includegraphics[scale=0.6]{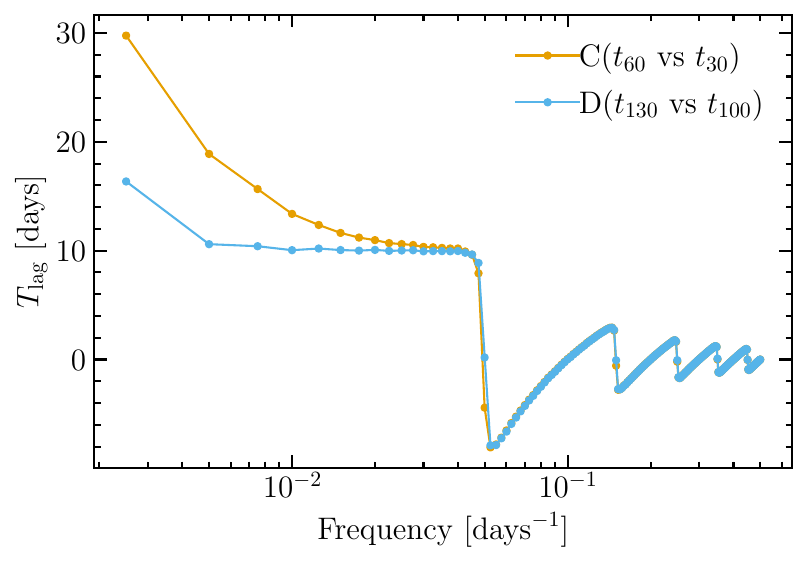}
\label{fig:DRW11(b)}}
\caption{(a) The DRW light curves for the damping timescale $t_{\mathrm{damp}}=30$ days, $60$ days, $100$ days, and $130$ days, respectively. All light curves are generated with the same random state; hence, these light curves are strongly correlated. The light curves with $t_{\mathrm{damp}}=30$ days or $100$ days are shifted by 10 days (see text). (b) The 1000-times averaged frequency-resolved lags in two cases, C($t_{60}$ vs $t_{30}$) and D($t_{130}$ vs $t_{100}$).}
\label{fig:DRW111}
\end{figure}

The thermal timescale ($\tau_{\mathrm{TH}} \sim 1/(\alpha \Omega_{\mathrm{K}})$) in the CHAR model represents the gas temperature fluctuation timescale, which increases with increasing radius and black-hole mass. Hence, different sources can have different characteristic variability timescales. We find that the variability timescale can also affect the frequency-resolved time lags. To illustrate this effect, we take the Damped Random Walk (hereafter DRW) model as an example. The light-curve generating algorithm is from AstroML Python package\footnote{\url{https://www.astroml.org/modules/generated/astroML.time_series.generate_damped_RW}}. One parameter of the DRW model is the damping timescale $\tau_{\mathrm{DRW}}$. We generate 410-day-long DRW signals in the same random state with four damping timescales of 30 days, 60 days, 100 days, and 130 days, respectively. Here, the damping timescales of 30 and 60 days represent the variability timescales of two light curves produced by a mock source, while 100 days and 130 days are for the other more luminous mock source. The cadence of DRW signals is 1 day. These DRW signals are as a function of time ($t$; in unit of days) denoted by $y_{30}$, $y_{60}$, $y_{100}$, and $y_{130}$, respectively. Then, we artificially shift the light curves $y_{30}(t)$ and $y_{100}(t)$ by $10$ days and obtain two new light curves, i.e., $y^{'}_{30}(t)=y_{30}(t+10)$ and $y^{'}_{100}(t)=y_{100}(t+10)$. We calculate frequency-resolved lags between 400-day-long signals $y_{60}(t)$ and $y_{30}(t+10)$ (hereafter case C) or between $y_{130}(t)$ and $y_{100}(t+10)$ (hereafter case D). The simulation is repeated 1000 times, and the average lags as a function of $f$ are shown in Fig.~\ref{fig:DRW11(b)}. It shows that at a fixed time span, the larger damping timescales result in smaller lags at the low-frequency regimes. Hence, this effect may also explain the reduced time lags in high luminosity sources.

\section{The frequency-resolved lags uncertainty of the model and observation}
In the Fourier analysis, the uncertainties of the time lag at low and high frequencies are large. The former is due to the small number of low-frequency signals, while the latter is due to the significant influence of signal noise. Hence, the uncertainty of observation is related to the time duration and the signal-to-noise ratio. In our CHAR simulations for NGC 5548, Mrk 335, and Mrk 817, we use the same time duration as observations but do not consider the measurement errors of the light curves.

\bibliographystyle{aasjournal}
\bibliography{export-bibtex.bib}

\end{document}